\documentclass[twocolumn,trackchanges]{aastex63}
\hypersetup{linkcolor=red,citecolor=blue,filecolor=cyan,urlcolor=magenta}

\usepackage{enumerate}
\usepackage{amsmath}
\usepackage{tikz}
\usetikzlibrary{shapes.geometric, arrows, positioning, fit}

\received{2024 May 1}
\revised{2025 January 12}
\accepted{2025 January 17}
\submitjournal{ApJ}

\shorttitle{Observing SN Neutrinos. V. Distance Estimation}
\shortauthors{Suwa et al.}
\graphicspath{{./}{fig/}}

\begin{document}

\title{Observing Supernova Neutrino Light Curves with Super-Kamiokande.\\ 
V. Distance Estimation with Neutrinos}

\correspondingauthor{Yudai Suwa}
\email{suwa@yukawa.kyoto-u.ac.jp}

\author[0000-0002-7443-2215]{Yudai Suwa}
\affiliation{Department of Earth Science and Astronomy, The University of Tokyo, Tokyo 153-8902, Japan}
\affiliation{Center for Gravitational Physics and Quantum Information, Yukawa Institute for Theoretical Physics, Kyoto University, Kyoto 606-8502, Japan}

\author[0000-0003-1409-0695]{Akira Harada}
\affiliation{Interdisciplinary Theoretical and Mathematical Sciences Program (iTHEMS), RIKEN, Wako, Saitama 351-0198, Japan}
\affiliation{National Institute of Technology, Ibaraki College, Hitachinaka 312-8508, Japan}

\author[0000-0002-0827-9152]{Masamitsu Mori}
\affiliation{Division of Science, National Astronomical Observatory of Japan, 2-21-1 Osawa, Mitaka, Tokyo 181-8588, Japan}

\author[0000-0001-6330-1685]{Ken'ichiro Nakazato}
\affiliation{Faculty of Arts and Science, Kyushu University, Fukuoka 819-0395, Japan}

\author[0000-0002-9234-813X]{Ryuichiro Akaho}
\affiliation{Faculty of Science and Engineering, Waseda University, Tokyo 169-8555, Japan}

\author[0000-0003-3273-946X]{Masayuki Harada}
\affiliation{Department of Physics, Okayama University, Okayama 700-8530, Japan}
\affiliation{Kamioka Observatory, Institute for Cosmic Ray Research, The University of Tokyo, Gifu 506-1205, Japan}

\author[0000-0003-0437-8505]{Yusuke Koshio}
\affiliation{Department of Physics, Okayama University, Okayama 700-8530, Japan}
\affiliation{Kavli Institute for the Physics and Mathematics of the Universe (Kavli IPMU, WPI), Todai Institutes for Advanced Study, \\ The University of Tokyo, Kashiwa 277-8583, Japan}

\author[0000-0003-4408-6929]{Fumi Nakanishi}
\affiliation{Department of Physics, Okayama University, Okayama 700-8530, Japan}

\author[0000-0002-9224-9449]{Kohsuke Sumiyoshi}
\affiliation{National Institute of Technology, Numazu College, Numazu 410-8501, Japan}

\author{Roger A. Wendell}
\affiliation{Department of Physics, Kyoto University, Kyoto 606-8502, Japan}
\affiliation{Kavli Institute for the Physics and Mathematics of the Universe (Kavli IPMU, WPI), Todai Institutes for Advanced Study, \\ The University of Tokyo, Kashiwa 277-8583, Japan}




\begin{abstract}

Neutrinos are pivotal signals in multi-messenger observations of supernovae (SNe). Recent advancements in the analysis method of supernova (SN) neutrinos, especially in quantitative analysis, have significantly broadened scientific possibilities. This study demonstrates the feasibility of estimating distances to SNe using neutrinos. This estimation utilizes the direct relationship between the radius of a neutron star (NS) and the distance to the SN, which is analogous to main-sequence fitting. The radius of an NS is determined with an approximate uncertainty of 10\% through observations such as X-rays and gravitational waves. By integrating this information, the distance to the SN can be estimated with an uncertainty of within 15\% at a 95\% confidence level. It has been established that neutrinos can pinpoint the direction of SNe, and when combined with distance estimates, three-dimensional localization becomes achievable. This capability is vital for follow-up observations using multi-messenger approaches. Moreover, more precise distance determinations to SNe through follow-up observations, such as optical observations, allow for accurate measurements of NS radii. This data, via the NS mass-radius relationship, could provide various insights into nuclear physics.

\end{abstract}



\section{Introduction} 
\label{sec:intro}

A supernova (SN) produces a large number ($\sim 10^{58}$) of neutrinos. These neutrinos, originating from thermal processes, are emitted almost isotropically. As a result, neutrinos generated in nearby supernovae (SNe) ensure their detectability \citep[see][and references therein]{2017hsn..book.1575J,  2018JPhG...45d3002H, 2019ARNPS..69..253M, 2023MNRAS.526.5900V,2024pjab.100.015}. It is noteworthy that neutrinos are emitted before electromagnetic radiation, as their production occurs while the shock wave is still confined within the star \citep{2013ApJ...778...81K,2016APh....81...39A}. Therefore, observing SN neutrinos is an initial step for further multiband observational studies that utilize the broad electromagnetic wavelength, extending from radio to gamma rays. In multi-messenger studies of SN research, neutrinos are regarded as extremely important from various scientific perspectives.

Recently, advances have been made in the methods used for the quantitative analysis of neutrinos from SNe. This field particularly focuses on the period starting a few seconds after the formation of the protoneutron star (PNS) from the SN. At this point, the PNS is no longer contracting,\footnote{This implies that the radius of the PNS has already converged to that of the cold neutron star (NS). In the subsequent discussion, the radius of the PNS and that of NS are not distinguished.} the falling back accretion onto it from the ejecta has become minor, and neutrinos start to come out from the deep inside of the PNS in a simpler way, through diffusion. This makes the situation much easier to describe physically, and quantitative analysis becomes possible. We mainly aim to understand this stage with long-term simulations \citep{2019ApJ...881..139S,2021PTEP.2021b3E01M,2022ApJ...925...98N}, find analytic ways to describe how neutrinos are emitted \citep{2021PTEP.2021a3E01S}, and construct the pipeline code for data analysis \citep{2022ApJ...934...15S, 2023ApJ...954...52H}.

In this paper, we investigate the potential of determining the distance to an SN through a quantitative analysis of SN neutrinos, emphasizing the straightforwardly derived relationship between the SN distance and the radius of the PNS. By determining the PNS radius, we can estimate the SN distance, and vice versa. Based on this relationship, we introduce a neutrino-based distance estimation method analogous to traditional astronomical techniques such as main-sequence fitting or spectroscopic parallax. In this method, the distance of a star is determined by comparing its absolute magnitude, derived from its position on the Hertzsprung-Russell diagram, with its observed magnitude. Similarly, our approach utilizes the thermal distribution of neutrinos, applying the Stefan-Boltzmann law and the radius of the neutron star--obtained from other observations like X-rays--to estimate distances. Thus, the simplicity of Eqs. \eqref{eq:D} and \eqref{eq:R} provides a straightforward quantitative framework for neutrino-based SN distance determination, analogous to how the main-sequence fitting relies on well-characterized stellar properties.

This paper is organized as follows. Section \ref{sec:mock} describes the model used to generate mock data, while Section \ref{sec:chi2} details the analysis of these mock samples for parameter estimation. Section \ref{sec:implication} discusses the implications of the findings, and Section \ref{sec:summary} provides a summary of the main results.

\section{Mock sampling}
\label{sec:mock}

In this work, we use the same method to perform parameter estimation as \cite{2022ApJ...934...15S} and \cite{2023ApJ...954...52H}, in which we use the solution for the neutrino light curve derived in \citet{2021PTEP.2021a3E01S}. The time evolution of the event rate and positron average energy are given by analytic functions of time. 
The parameter dependence on the mass and radius of the PNS are also explicitly presented.

The event rate is given by
\begin{align}
\mathcal{R}&
=720\,{\rm s^{-1}}
\left(\frac{M_{\rm det}}{32.5\,{\rm kton}}\right)
\left(\frac{D}{10\,{\rm kpc}}\right)^{-2}
\nonumber\\
&\times
\left(\frac{M_{\rm PNS}}{1.4\,M_\odot}\right)^{15/2}
\left(\frac{R_{\rm PNS}}{10\,{\rm km}}\right)^{-8}
\left(\frac{g\beta}{3}\right)^{5}\nonumber\\
&\times\left(\frac{t+t_0}{100\,{\rm s}}\right)^{-15/2},
\label{eq:detection_rate}
\end{align}
where 
$M_{\rm det}$ is the detector mass with 32.5~kton corresponding to the entire inner detector volume of Super-Kamiokande (SK),\footnote{In this study, we employ the full 32.5 kton volume of the SK inner detector. This is because, at least for a Galactic SN, the timescale of data analysis is short, and we can avoid significant contamination from backgrounds \citep{2022ApJ...938...35M}. For a more detailed discussion, see \cite{2019ApJ...881..139S}. } $D$ is the distance between the SN and Earth, $M_{\rm PNS}$ is the PNS mass,  $R_{\rm PNS}$ is the PNS radius,\footnote{This corresponds to the radius after PNS contraction by neutrino cooling.} $g$ is the surface structure correction factor, and $\beta$ is the opacity boosting factor from coherent scattering \citep[see][for details]{2021PTEP.2021a3E01S}. 
The timescale $t_0$ is given by
\begin{align}
t_0&=210\,{\rm s}
\left(\frac{M_{\rm PNS}}{1.4\,M_\odot}\right)^{6/5}
\left(\frac{R_{\rm PNS}}{10\,{\rm km}}\right)^{-6/5}\nonumber\\
&\times\left(\frac{g\beta}{3}\right)^{4/5}
\left(\frac{E_{\rm tot}}{10^{52}\,{\rm erg}}\right)^{-1/5},
\label{eq:t0}
\end{align}
where $E_{\rm tot}$ is the total energy emitted by all flavors of neutrinos.
By integrating Eq. \eqref{eq:detection_rate} with Eq. \eqref{eq:t0}, 
the expected total number of events is
\begin{align}
    N&=\int_0^{\infty} \mathcal{R}(t)dt\nonumber\\
    &=89\left(\frac{M_{\rm det}}{32.5\,{\rm kton}}\right)
\left(\frac{D}{10\,{\rm kpc}}\right)^{-2}
\left(\frac{M_{\rm PNS}}{1.4\,M_\odot}\right)^{-3/10}\nonumber\\
&\times
\left(\frac{R_{\rm PNS}}{10\,{\rm km}}\right)^{-1/5}
\left(\frac{g\beta}{3}\right)^{-1/5}
\left(\frac{E_{\rm tot}}{10^{52}\,{\rm erg}}\right)^{13/10}.
\label{eq:total_number}
\end{align}
For the canonical parameters used in this paper ($M_{\rm det}=32.5$ kton, $D=8$ kpc, $M_{\rm PNS}=1.52 M_\odot$, $R_{\rm PNS}=12.4$ km, $g\beta=1.6$, and $E_{\rm tot}=10^{53}$ erg), the expectation number becomes $N=2940$.

The average energy of created positrons is given by
\begin{align}
E_{e^+}
&=25.3\, {\rm MeV}
\left(\frac{M_{\rm PNS}}{1.4\,M_\odot}\right)^{3/2}\nonumber\\
&\times
\left(\frac{R_{\rm PNS}}{10\,{\rm km}}\right)^{-2}
\left(\frac{g\beta}{3}\right)
\left(\frac{t+t_0}{100\,{\rm s}}\right)^{-3/2}.
\label{eq:detection_energy}
\end{align}
For the energy distribution, we employ the Fermi-Dirac distribution function for neutrinos, which allows us to calculate the distribution of the positron. 
Note that the above estimates are based on the simple expression for the cross section of the inverse beta decay. More precise expressions are given in \cite{1999PhRvD..60e3003V,2003PhLB..564...42S}, which we used for our numerical estimates in \cite{2022ApJ...925...98N}.

Importantly, by combining Eqs. \eqref{eq:detection_rate} and \eqref{eq:detection_energy} we find
\begin{align}
D=&
10\,{\rm kpc}
\left(\frac{\mathcal{R}}{720\,{\rm s^{-1}}}\right)^{-1/2}
\left(\frac{E_{e^+}}{25.3\, {\rm MeV}}\right)^{5/2}\nonumber \\
&\times \left(\frac{M_{\rm det}}{32.5\,{\rm kton}}\right)^{1/2}
\left(\frac{R_{\rm PNS}}{10\,{\rm km}}\right).
\label{eq:D}
\end{align}
Note that it is independent of $t$, $g$, and $\beta$. This equation has two meanings. If we know the radius $R_{\rm PNS}$, we can estimate the distance $D$  and vice versa. In \cite{2022ApJ...934...15S, 2023ApJ...954...52H}, we assumed that $D$ is known by other means so that we could constrain the radius. In the following, we will show how accurately we can estimate the distance before the other observations are available.

Based on the equations presented here, we perform 100 Monte Carlo simulations of SN neutrinos, each using different random seeds \citep[see Figure 1 in][for a specific example]{2022ApJ...934...15S}. In the next section, we explain the data analysis method.

\section{Parameter estimation}
\label{sec:chi2}

Parameter estimation is conducted by fitting the Monte-Carlo data described in Section \ref{sec:mock} to analytic solutions, aiming to assess the accuracy and reliability of our method \citep[see Figure 2 in][]{2022ApJ...934...15S}. Detailed descriptions of the numerical setting, the fitting procedure, and the statistical methods used for the analysis are provided below.

Here, we employ the Gaussian-approximation analysis \citep[see][for details]{2023ApJ...954...52H}, i.e., the chi-square fitting using the event rate and the average energy. This is because we are interested in Galactic SNe, so the expected event number is large enough for the Poisson distribution to be approximated by the Gaussian distribution. Also, in this work, we increase the number of parameters from three to four so that computational cost becomes greater than the previous work and spectral likelihood analysis becomes time-consuming. For binning and the probability density function definition, see \cite{2022ApJ...934...15S}. For completeness, we summarize the following.
The time bins are calculated by the equations:
\begin{align}
    t_i&=t_{i-1} + \Delta t_i,\\
    \Delta t_i&= A\Delta t_{i-1},
\end{align}
where $\Delta t_i$ represents the time width of the $i$-th time bin, and $A$ is a constant. This constant is determined using the first time bin, $t_1$, and the last time bin of the analysis, $t_{\rm end}$. For this paper, we set $\Delta t_1$ to 0.5 and $t_{\rm end}$ to 100 seconds. We chose the number of bins, $N$, to be 20, resulting in an approximate value of $A$ equal to 1.206.
The $\chi^2$ is calculated as follows:
\begin{align}
    \chi^2=\sum_{i=1}^N \frac{(O_i-X_i)^2}{\sigma_i^2},
    \label{eq:chi2}
\end{align}
where $O_i$, $X_i$, and $\sigma_i$ represent the observed value, the expected value, and the variance, respectively, each associated with the time index $i$. For the event rate ($X=\mathcal{R}$), the variance is calculated using $\sigma_i^2=\mathcal{R}_i^2/N_i$, where $N_i$ denotes the number of events in the $i$-th bin. For the average energy ($X=E_{e^+}$), we adopt $\sigma_i^2=(0.05E_{e^+})^2$, as outlined by \cite{2022ApJ...925...98N}, which demonstrates that the statistical error of the average energy is at the level of several percent.

In \citet{2022ApJ...934...15S}, we employed the joint probability density function (PDF) for the measured parameters as
\begin{align}
    \mathcal{P}(M_{\rm PNS},R_{\rm PNS}, E_{\rm tot})\propto e^{-\chi^2(M_{\rm PNS},R_{\rm PNS}, E_{\rm tot})/2}.
    \label{eq:PDF}
\end{align}
Instead of Eq. \eqref{eq:PDF} that assumes the uniform prior for all parameters, we employ the following PDF, including Gaussian prior for the radius as
\begin{align}
&\mathcal{P}(M_{\rm PNS},R_{\rm PNS}, E_{\rm tot},D)  \nonumber\\
&\propto e^{-\chi^2(M_{\rm PNS},R_{\rm PNS}, E_{\rm tot},D)/2}  
\times e^{-(R_{\rm PNS}-\bar R)^2/(2\sigma_R^2)},
    \label{eq:PDF2}
\end{align}
where $\bar R$ and $\sigma_R$ are the expected mean value and its uncertainties of PNS radius. In this work, we employ $\bar R=12.4$ km and $\sigma_R=0.7$ km based on \cite{2021ApJ...918L..28M}, suggesting that the NS radius is not strongly dependent on its mass.
For the other parameters, we employ uniform prior within $1.0<M_{\rm PNS}/M_\odot<2.0$, $0.5<E_{\rm tot}/(10^{53}\,{\rm erg})<2.0$, and $3<D/{\rm kpc}<13$, respectively.

\begin{figure}[tbp]
\centering
\includegraphics[width=0.45\textwidth]{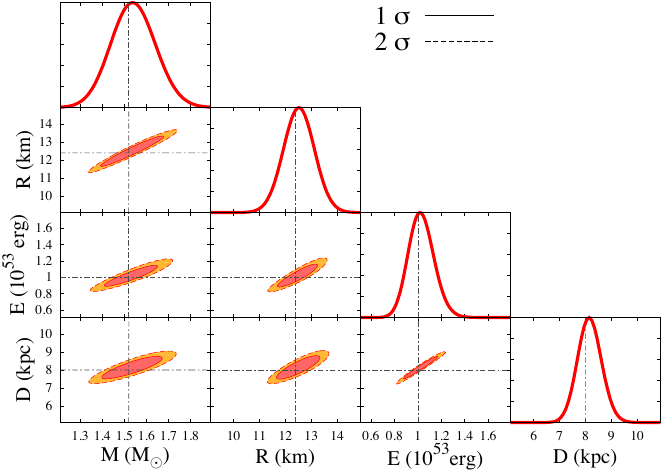}
\caption{A sample of probability density function (PDF) determined by Eq. \eqref{eq:PDF2}.
Contours with solid and dashed lines correspond to $\mathcal{P}/\mathcal{P}_{\rm max}=1/e$ (0.368, corresponding to $1\sigma$) and $1/e^2$ (0.135, $2\sigma$), respectively, where $\mathcal{P}_{\rm max}$ is the maximum value of the PDF. 
}
\label{fig:chi2}
\end{figure}

Figure \ref{fig:chi2} presents the distribution of $\mathcal{P}$. Solid and dashed contours in this figure represent $\mathcal{P}/\mathcal{P}_{\rm max}=1/e$ (equivalent to 0.368 or 1$\sigma$) and $1/e^2$ (0.135, corresponding to 2$\sigma$), respectively, with $\mathcal{P}_{\rm max}$ denoting the peak value of $\mathcal{P}$. It is important to note that the uncertainties depicted are based on a single realization. Given that the observed data can exhibit variations due to Poisson statistics, conducting Monte Carlo simulations for multiple realizations is crucial to accurately assess the expected parameter sensitivity (or expected error) in preparation for actual observational data.

\begin{table}[bp]
    \centering
    \caption{Expected Values and Statistical Errors with only Neutrinos}
    \begin{tabular}{ccccc}
        & input &  Median & 68\% & 95\% \\
    \hline
     $M_{\rm PNS}$ ($M_\odot$) & 1.52 & 1.58 & $^{+0.13}_{-0.12}$ & $^{+0.26}_{-0.24}$\\
     $R_{\rm PNS}$ (km) &  12.4 & 12.5 & $^{+0.7}_{-0.7}$ & $^{+1.4}_{-1.4}$\\
     $E_{\rm tot}$ ($10^{53}$ erg) & 1.00 & $1.05$ & $^{+0.15}_{-0.13}$  &  $^{+0.31}_{-0.25}$\\ 
     $D$ (kpc) &  $8.00$ & 8.10 & $^{+0.60}_{-0.56}$  &  $^{+1.24}_{-1.08}$\\ 
    \hline
    \end{tabular}
    \label{tab:table}
\end{table}

The expected parameter sensitivity is assessed through 100 realizations of the aforementioned model.\footnote{The calculation with 1,000 realizations yields almost identical results, indicating that the calculation has reasonably converged \citep[see][]{2022ApJ...934...15S}.} Each realization, processed via Monte Carlo simulations, yields a variety of best-fit values in accordance with Poisson statistics. However, the cumulative average demonstrates that the preset input values are most likely to be accurate, with a significant decrease in probability density for values diverging from the initial inputs. To estimate the expected parameter sensitivity, we use the median of the compiled average PDF. The uncertainty levels of 68\% and 95\% are determined by the range of parameters that correspond to these specific probability levels, centered around the median. The findings are compiled in Table \ref{tab:table}.
Due to the uncertainty imposed by the priors on the radius, it is evident that the precision in determining the mass and total energy is subject to greater uncertainty compared to our previous study \citep[see Table 1 in][]{2022ApJ...934...15S}. However, it is noteworthy that the distance to the SN has been determined with a precision of within 15\% at a 95\% confidence level, representing a significant new piece of information.
To assess the impact of the number of detected neutrinos on uncertainty, we repeated the procedure for a case where $D=5$ kpc. The uncertainty in this case is 14\%, indicating that the uncertainty is primarily determined by the prior imposed on $R$. 
We also investigate the impact of prior distribution, assuming $\sigma_R=1$ km. The resulting error in the distance estimation is approximately 20\% at a 95\% confidence level.

\section{Implication}
\label{sec:implication}

Here, we discuss the application of distance estimation. 
The flowchart of the analysis is shown in Figure \ref{fig:flowchart}.

\tikzstyle{startstop} = [rectangle, rounded corners, minimum width=2cm, minimum height=1cm,text centered, fill=red!20]
\tikzstyle{process} = [rectangle, minimum width=2cm, minimum height=1cm, text centered, text width=3.3cm, fill=orange!20]
\tikzstyle{arrow} = [thick,->,>=stealth]
\tikzstyle{label} = [text width=4cm, text centered]

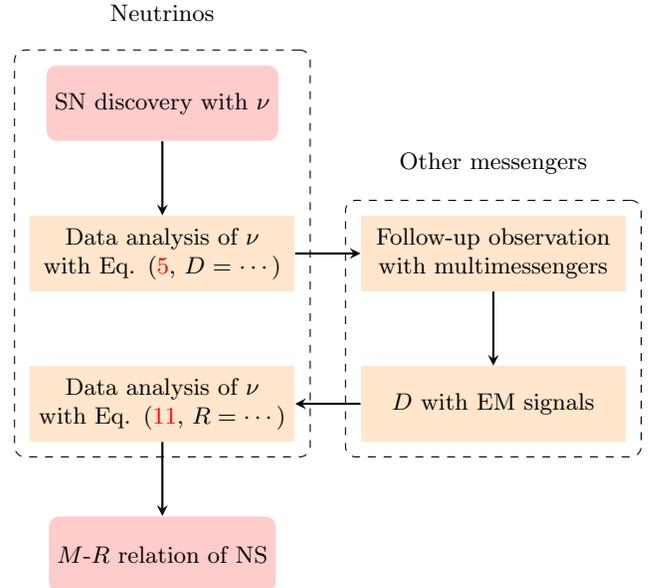
\begin{figure}[tbp]
\centering
\begin{tikzpicture}[node distance=2cm]

\node (start) [startstop] {SN discovery with $\nu$};
\node (process1) [process, below of=start] {Data analysis of $\nu$ with Eq. (\ref{eq:D}, $D=\cdots$)};
\node (process2) [process, right of=process1, xshift=2.4cm] {Follow-up observation with multimessengers};
\node (process3) [process, below of=process2] {$D$ with EM signals};
\node (process4) [process, below of=process1] {Data analysis of $\nu$ with Eq. (\ref{eq:R}, $R=\cdots$)};
\node (stop) [startstop, below of=process4] {$M$-$R$ relation of NS};

\node (fit1) [fit=(start) (process1) (process4), draw, dashed, inner sep=0.2cm, rectangle, rounded corners] {};
\node [above of=fit1, node distance=3.2cm] {Neutrinos};

\node (fit2) [fit=(process2) (process3), draw, dashed, inner sep=0.2cm, rectangle, rounded corners] {};
\node [above of=fit2, node distance=2.2cm] {Other messengers};

\draw [arrow] (start) -- (process1);
\draw [arrow] (process1) -- ++(2,0) |- (process2);
\draw [arrow] (process2) -- (process3);
\draw [arrow] (process3) -- ++(-2,0) |- (process4);
\draw [arrow] (process4) -- (stop);

\end{tikzpicture}
\caption{Flowchart depicting the process of supernova (SN) discovery using neutrinos ($\nu$) and subsequent analyses. The chart outlines steps from initial detection through data analysis involving specific equations, integration of multimessenger follow-up observations, distance measurement with electromagnetic (EM) signals, and culminating in the determination of the mass-radius ($M$-$R$) relationship of neutron stars (NS).}
\label{fig:flowchart}
\end{figure}

\begin{figure}[tbp]
\centering
\includegraphics[width=0.45\textwidth]{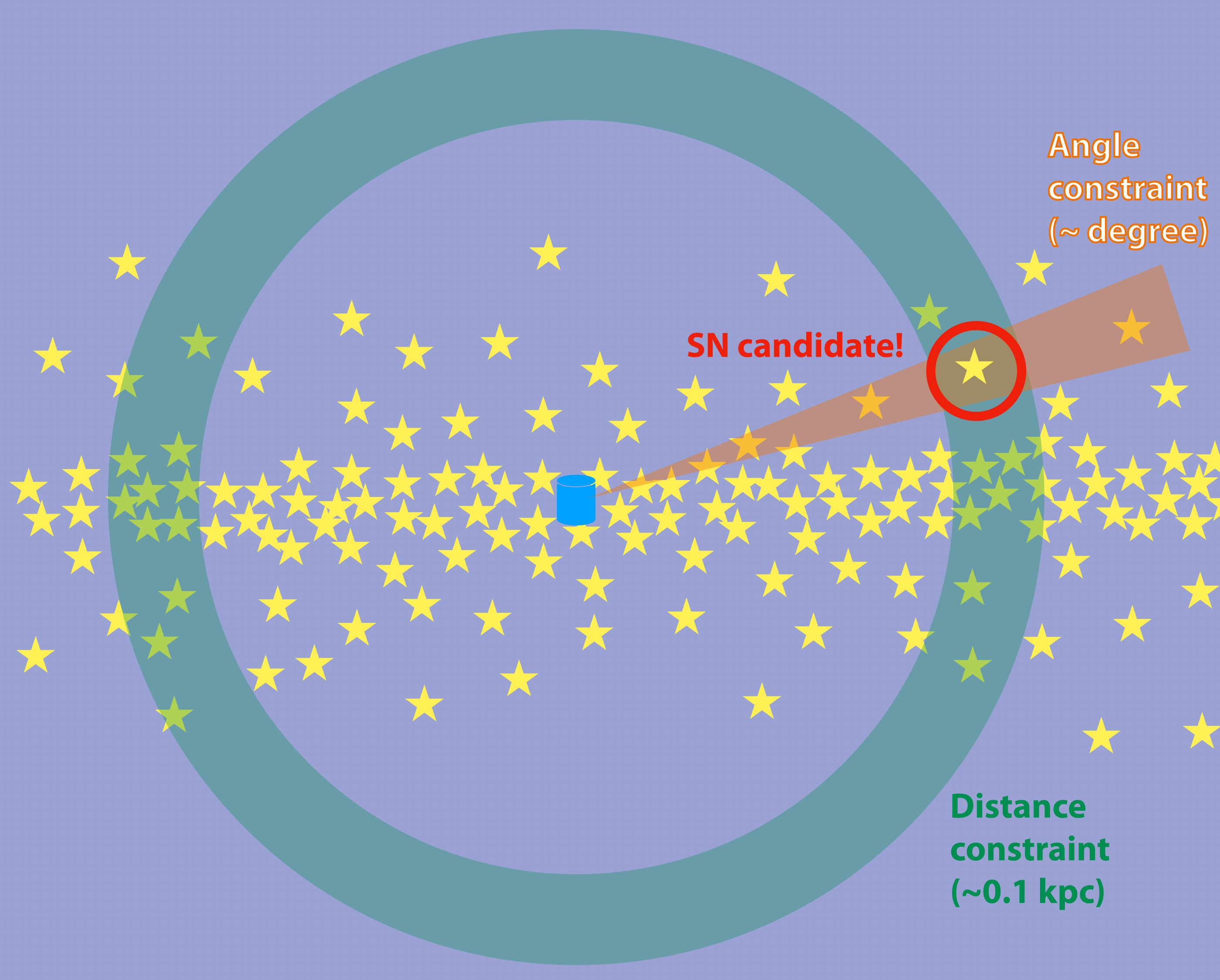}
\caption{A schematic image for identifying the supernova candidate based on the neutrino signals detected at SK (blue cylinder at the center). Analysis of the signals gives directional information (orange triangle region) and distance measurement (green circle). Combining them, the three-dimensional position of the supernova progenitor may be identified.
}
\label{fig:schematic_position}
\end{figure}

After observing neutrinos, it takes time for the shock wave to propagate through the star. Hence, there is a delay before the onset of electromagnetic radiation: approximately $10^5$ seconds for red supergiants, $10^4$ seconds for blue supergiants, and $10^2$ seconds for Wolf-Rayet stars \citep{2013ApJ...778...81K}. On the other hand, in principle, neutrino detectors can issue an alert within a few minutes \citep{2016APh....81...39A,2021NJPh...23c1201A,2024arXiv240306760K}.
If the SN progenitor can be identified through neutrino observations, conducting multi-messenger observations of the shock breakout becomes possible. Previously, direction determination by neutrinos has been discussed in various studies \citep{1999PhRvD..60c3007B,2002PThPh.107..957A,2016APh....81...39A,2019PhRvD.100j3005L,2020ApJ...899..153M, 2024arXiv240306760K}, in which the direction would be determined within several degrees. On the other hand, this study has revealed that it is possible to estimate distances using neutrinos alone (Eq. \ref{eq:D}) \citep[see][for a different approach]{2021arXiv210110624S, 2024JCAP...02..008B}. Combining the estimated direction and distance makes it possible to determine the three-dimensional position of the SN progenitor (see Figure \ref{fig:schematic_position}). If there is only one massive star at the estimated position, it becomes possible to uniquely determine the progenitor star before electromagnetic observation of the explosion.\footnote{A list of nearby massive stars that may produce supernovae has been compiled in \cite{2024MNRAS.529.3630H}, and matching the three-dimensional positions provided by neutrinos with this list should allow us to identify the progenitor star of the supernova.} This information is essential, particularly for follow-up observations by telescopes with limited sky coverage. At this stage, the accuracy of distance estimation depends on the accuracy of determining the NS radius, which is $\mathcal{O}$(10)\%.

Next, we consider the use of neutrino data after the realization of electromagnetic observations. Suppose that the distance can be determined with an accuracy of $\mathcal{O}$(1)\% through observations by, for instance, optical telescopes.\footnote{When considering parallax measurements with Gaia, bright objects can achieve distance estimates with approximately 1\% precision, even at distances exceeding 1 kpc \citep{2023A&A...674A...1G}.} In that case, imposing a prior distribution on the distance and estimating the NS radius is possible. Here, Eq. \eqref{eq:D} should be changed as 
\begin{align}
R_{\rm PNS}=&
10\,{\rm km}
\left(\frac{\mathcal{R}}{720\,{\rm s^{-1}}}\right)^{1/2}
\left(\frac{E_{e^+}}{25.3\, {\rm MeV}}\right)^{-5/2}\nonumber \\
&\times \left(\frac{M_{\rm det}}{32.5\,{\rm kton}}\right)^{-1/2}
\left(\frac{D}{10\,{\rm kpc}}\right).
\label{eq:R}
\end{align}
Furthermore, as demonstrated in previous studies \citep{2022ApJ...934...15S,2023ApJ...954...52H}, it is also possible to independently estimate the mass of the PNS. Combining these makes it possible to constrain the mass-radius relationship of NSs from neutrino observations. For instance, if the distance to an SN can be determined with 1\% accuracy through optical observations, the precision in determining the NS radius would also approximate $\sim 1$\% for a nearby SN. This represents a tenfold improvement in precision compared to the most stringent current observational limits, significantly enhancing the constraints on nuclear physics.

\section{Summary and discussion}
\label{sec:summary}

In this study, we show that the quantitative analysis of neutrinos, which has recently become possible, can independently estimate the distance to a supernova explosion with an accuracy of 10\%. This methodology relies on prior information about the neutron star radius derived from supplementary observations. When combined with neutrino-based determinations of the supernova's direction, this approach enables three-dimensional localization, which is crucial for follow-up observations. Moreover, if the supernova's distance is further refined through electromagnetic observations, this enhanced distance accuracy can reciprocally refine parameter estimations, thereby enabling a highly precise determination of neutron star radii.

We simplify the analysis by focusing on late-phase neutrino emission. Specifically, neutrinos are emitted through a quasi-static diffusion process within a time-stationary neutron star density distribution. As demonstrated by \cite{2021PTEP.2021a3E01S}, neutrino emission can be expressed analytically, and this study is based on that analytic formula. However, additional physical factors that may influence neutrino emission should be accounted for as systematic errors. For instance, the effect of neutrino oscillations causes flavor transitions, while the neutrino spectrum becomes nearly flavor-independent at late times. These factors include uncertainties in numerical simulations of supernovae and detector response, which are essential for a more accurate evaluation. While these factors fall beyond the primary scope of this study, their inclusion in uncertainty evaluations will be essential for a more comprehensive analysis. Consequently, the uncertainties reported here reflect only statistical errors.
We are working on these aspects and systematic error evaluations, which will be reported in forthcoming papers.

\acknowledgments

YS thanks A. Tanikawa and H. Uchida for useful discussions. This work is supported by JSPS KAKENHI Grant Numbers JP19H05811, JP20H00174, JP20H01904, JP20H01905 JP20K03973,  JP20H04747, JP21K13913, JP23KJ2150, JP24H02236, JP24H02245, JP24K00632, JP24K00668, and JP24K07021.
Numerical computations were, in part, carried out on a computer cluster at CfCA of the National Astronomical Observatory of Japan.
This work was partly supported by MEXT as ``Program for Promoting Researches on the Supercomputer Fugaku'' (Toward a unified view of the universe: from large scale structures to planets). This work was partially carried out by the Inter-University Research Program of the Institute for Cosmic Ray Research (ICRR), the University of Tokyo.

\bibliographystyle{aasjournal}



\end{document}